\documentclass[traditabstract, letter]{aa}

\usepackage{graphicx}
\usepackage{txfonts}
\usepackage{natbib}
\bibpunct{(}{)}{;}{a}{}{,}
\begin{document}
\author{
  S. Heyminck\inst{1}\and
  U.U. Graf\inst{2}\and
  R. G\"usten\inst{1}\and
  J. Stutzki\inst{2}\and
  H.W. H\"ubers\inst{3,}\inst{4}\and
  P. Hartogh\inst{5}
}
\institute{
  Max-Planck-Institut f\"ur Radioastronomie, Auf dem H\"ugel
  69, 53121 Bonn / Germany
  \and
  I. Physikalisches Institut der Universit\"at zu K\"oln, Z\"ulpicher
  Stra{\ss}e 77, 50937 K\"oln / Germany
  \and
  Deutsches Zentrum f\"ur Luft- und Raumfahrt, Institut f\"ur Planetenforschung,
  Rutherfordstra{\ss}e 2, 12489 Berlin / Germany
  \and
  Institut f\"ur Optik und Atomare Physik, Technische Universit\"at Berlin, Hardenbergstra{\ss}e 36, 10623 Berlin / Germany
  \and
  Max-Planck-Institut f\"ur Sonnensystemforschung, Max-Planck-Stra{\ss}e 2,
  37191 Katlenburg-Lindau / Germany
}
\title{GREAT: the SOFIA high-frequency heterodyne instrument}
\date{Received / Accepted }
\abstract
{
  We describe the design and construction of GREAT, the
  {\underline{G}}erman {\underline{RE}}ceiver for
  {\underline{A}}stronomy at {\underline{T}}erahertz frequencies
  operated on the Stratospheric Observatory for Infrared Astronomy
  (SOFIA). GREAT is a modular dual-color heterodyne instrument for
  high-resolution far-infrared (FIR) spectroscopy. Selected for
  SOFIA's Early Science demonstration, the instrument has successfully
  performed three Short and more than a dozen Basic Science flights
  since first light was recorded on its April 1, 2011 commissioning
  flight.

  We report on the in-flight performance and operation of the receiver
  that -- in various flight configurations, with three different
  detector channels -- observed in several science-defined frequency
  windows between 1.25 and 2.5$\,$THz. The receiver optics was
  verified to be diffraction-limited as designed, with nominal
  efficiencies; receiver sensitivities are state-of-the-art, with
  excellent system stability. The modular design allows for the
  continuous integration of latest technologies; we briefly discuss
  additional channels under development and ongoing improvements for
  Cycle 1 observations.

  GREAT is a principal investigator instrument, developed by a
  consortium of four German research institutes, available to the
  SOFIA users on a collaborative basis.  } 

\titlerunning{GREAT: the SOFIA high frequency heterodyne instrument}

\maketitle

%
%
\section{Introduction}
\begin{figure*}[htbp]
 \unitlength1cm
 \begin{picture}(18.,10.)
  \put(0,0.3){\includegraphics[width=10.4cm, angle=0]{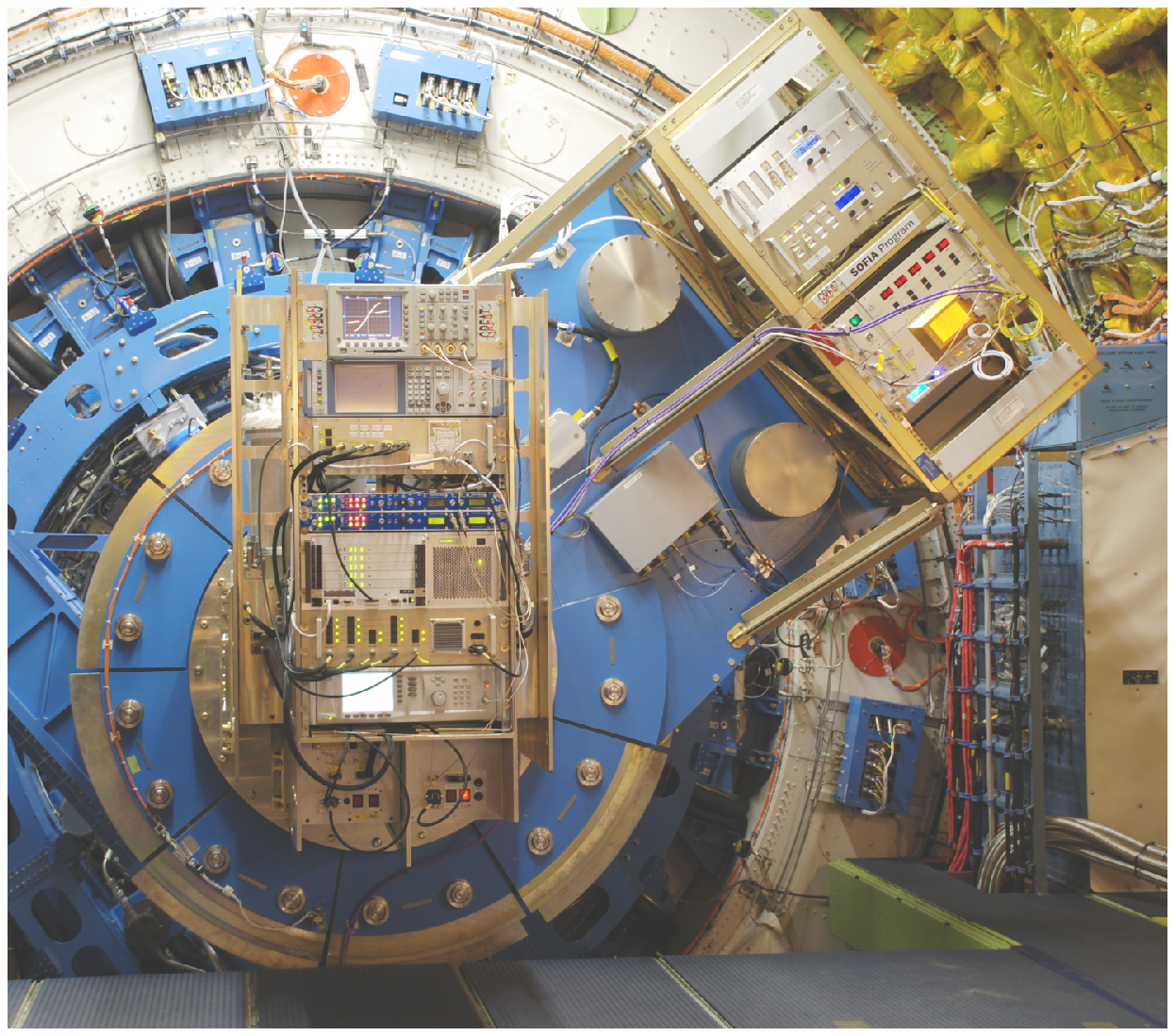}}
  \put(10.8,0.3){\includegraphics[width=7.2cm,angle=0]{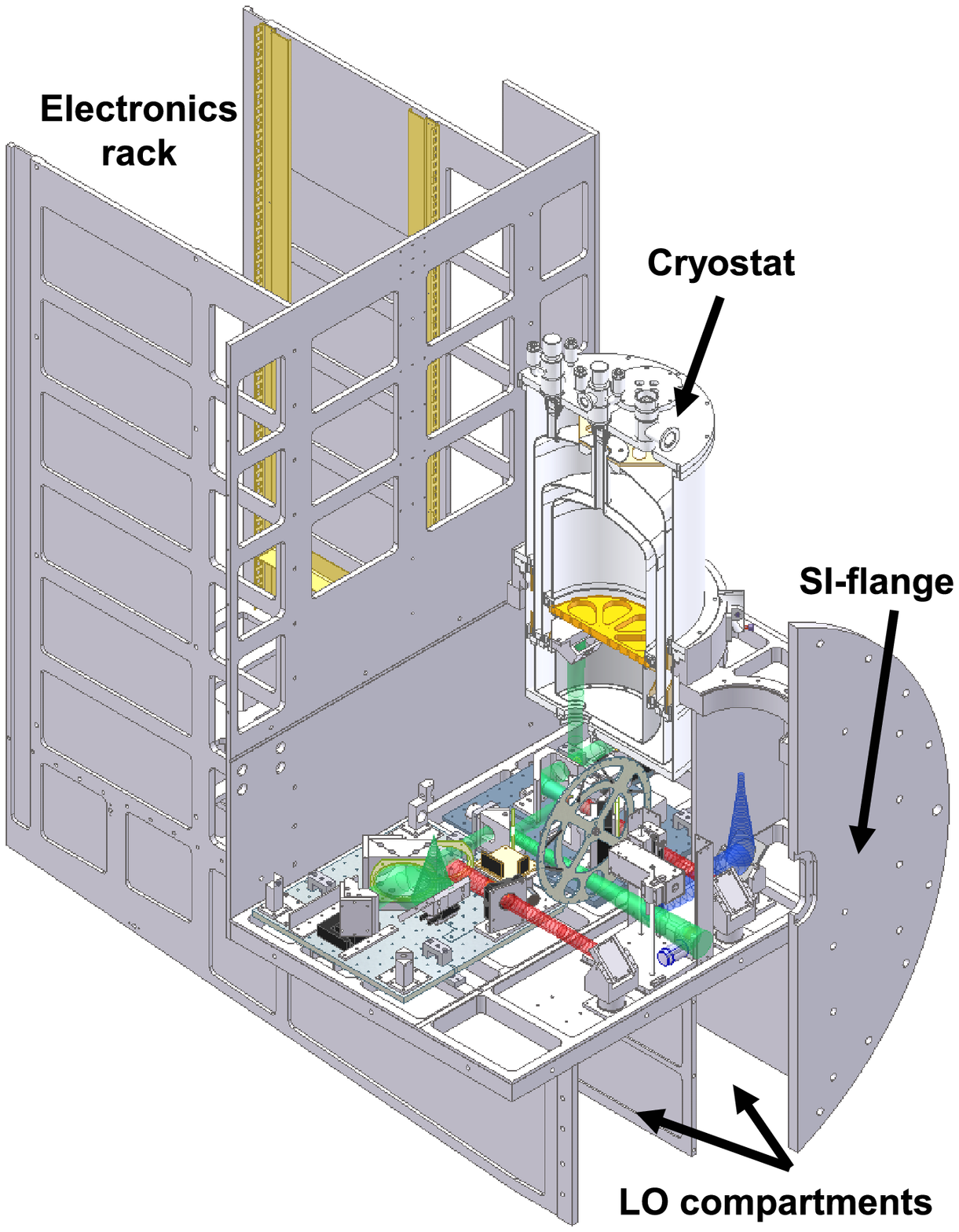}}
 \end{picture}
 \caption{
 \label{GREAT_flange}\label{GREAT_3d}
 Left: GREAT mounted to the instrument flange, in stow position (the
 telescope parks at $\rm 40\,^o$ elevation). During observations GREAT
 (together with the counterweight at top right) rotates $\rm
 \pm20\,^o$ from the vertical.  Right: Basic structural components of
 GREAT. One of the two cryostats, the location of the LO compartments
 and position of the main optics benches are shown. A more detailed
 view of the optical layout is given in Fig. \ref{GREAT_optics}.}
\end{figure*}

With SOFIA \citep{2009ASPC..417..101B, 2012Young} a new platform for
inter alia high-resolution far-infrared spectroscopy has begun science
operation.  Building on the legacy of the Kuiper Airborne Observatory
\citep{1981SPIE..265....1G}, SOFIA with its projected 20 years
operational lifetime will complement and carry on the science heritage
of Herschel \citep{2010A&A...518L...1P}. Since the first pioneering
high-resolution FIR spectrometers were flown on the KAO \citep[and
references therein]{1984abas.symp..320B, 1981SPIE..280..101P,
  1987IJIMW...8.1541R}, new technologies have enabled the development
of instruments with -- in those days -- unobtainable performances and
sensitivities. HIFI \citep{2010A&A...518L...6D}, the heterodyne
spectrometer currently flying on-board Herschel, and GREAT, the
subject of this paper, are the most advanced implementations for
actual science operation.

The development of GREAT 
\footnote{GREAT is a development by the MPI f\"ur Radioastronomie
  (Principal Investigator: R. G\"usten) and the KOSMA$/$Universit\"at
  zu K\"oln, in cooperation with the MPI f\"ur Sonnensystemforschung
  and the DLR Institut f\"ur Planetenforschung.}
, the {\underline{G}}erman {\underline{RE}}ceiver for
{\underline{A}}stronomy at {\underline{T}}erahertz frequencies with
SOFIA, initially begun in 2000, not least as pathfinder for the
Herschel satellite mission.

Owing to severe SOFIA project delays, it was not until October 2010,
however, that the instrument was shipped to the NASA Dryden Aircraft
Operations Facility (DAOF) in Palmdale, CA. On SOFIA's base there,
integration into the aircraft and on-ground commissioning proceeded
notably smooth, and first light was recorded already on April 1, 2011,
during SOFIA's observatory characterization flight OCF\#4. Since then,
GREAT has been operated for more than a dozen flights in SOFIA's Basic
Science program. As a PI-class instrument GREAT is operated by the
consortium only, but by agreement with the SOFIA director the
instrument has been made available to the SOFIA communities on a
collaborative basis.

In the following we focus mainly on the configuration and performance
of the instrument as flown during the Short and Basic Science periods
of SOFIA. A more detailed technical description of GREAT, including
ongoing developments, will be published elsewhere \citep[in
prep.]{heyminck2012}.

\begin{table}[ht]
  \caption{\label{GREAT_channels}Science Opportunities with GREAT}
  \begin{center}
    \begin{tabular}{|l|c|p{5.2cm}|}
      \hline
                 & RF tuning   & astrophysical lines                                      \\
                 &   [GHz]     &                                                          \\\hline\hline
      L1$\rm_a$  & 1252 - 1392 & {CO(11-10), CO(12-11), OD, SH, H$_2$D$^+$, HCN, HCO$^+$} \\
      L1$\rm_b$  & 1417 - 1520 & {[NII], $^{(13)}$CO(13-12), HCN, HCO$^+$}                \\\hline
      L2         & 1815 - 1910 & {NH$_3$(3-2), OH($^2\Pi_{1/2}$), CO(16-15), [CII]}       \\\hline
      M$\rm_a$   & 2507 - 2514 & OH($^2\Pi_{3/2}$)                                        \\
      M$\rm_b$   & 2670 - 2680 & HD(1-0)                                                  \\\hline
      H          & 4750 - 4770 & [OI]                                                     \\\hline
   \end{tabular}
 \end{center}
 Suffixes $\rm_a$ and $\rm_b$ denote the different LO-chains available for this receiver
 channel to extend their RF-coverage. It is not possible to swap these LO-chains during flight.

 Status: channels L1 $\&$ L2 were operated during Early and Basic Science, 
 M$\rm_a$ during Basic Science only; M$\rm_b$ and H are under development (see the following sections).
\end{table}

\begin{figure}[t]
  \begin{center}
    \includegraphics[width=9.cm,angle=0]{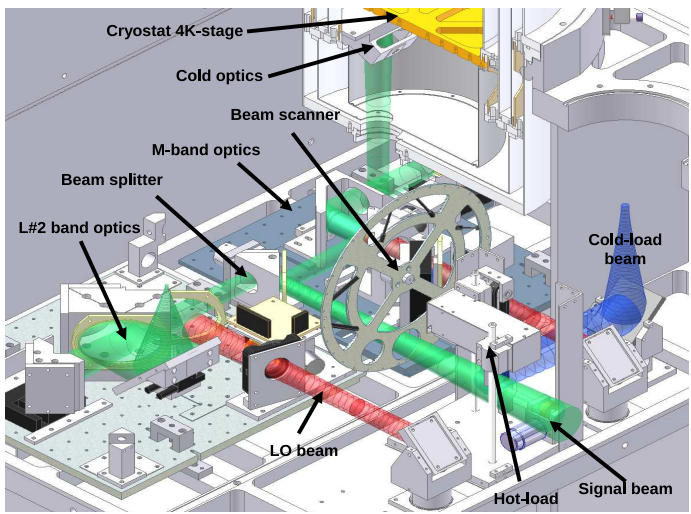}
  \end{center}
  \caption{ \label{GREAT_optics} Optics layout: A central polarizer
    inside the optics compartment splits and re-directs the incoming
    radiation to the two optics plates of the instrument. A common
    beam scanning mechanism (for internal alignment) and the
    calibration loads complete the optical system.}
\end{figure}

\section{Science requirements}
The users' desire for wide radio frequency (RF) coverage (1 --
5$\,$THz), with highest possible intermediate frequency (IF) bandwidth
($>3\,$GHz) and spectral resolution ($<100\,$kHz) is confronted with
current limitations of THz technologies. Despite impressive advances
during recent years, the operating bandwidth of the local oscillator
(LO) sources in particular is $\Delta\nu / \nu \sim 0.1$ only (and
significantly less above 2$\,$THz). Therefore, from the beginning
\citep{2000SPIE.4014...23G} GREAT was designed as a highly modular
instrument: within the framework of a common infrastructure a choice
of two frequency-specific channels can be operated simultaneously. For
Early Science three detector channels with a total of four LO sources
were flown (Table \ref{GREAT_channels}), two more are being developed.
The instrument configuration can be changed between flights, while
GREAT stays installed on SOFIA.

\section{Instrument description}
GREAT is composed of a wide variety of common modules out of which the
desired flight configuration set-up is formed together with the
frequency specific subunits \citep{2008SPIE.7014E..33H}.

The mechanical support structure of GREAT (consisting of lightweight,
certified aluminum Al-2024 plates) is connected to the science
instrument flange of the telescope (Fig. \ref{GREAT_flange}). This
frame carries the optics box with the common and channel-specific
optical components to guide the signals from sky and from the LOs to
the mixers. The receiver cryostats and the calibration unit are
clamped to the upper side of the structure with tight tolerance, which
allows a reproducible re-positioning of the units when the instrument
configuration is changed.

The structure carries the front-end control electronics on the forward
side and the two LOs underneath the optics box (Fig. \ref{GREAT_3d}).
The optics compartment represents the pressure boundary to the
aircraft, is open toward the telescope cavity and hence operates at
ambient pressure to minimize absorption losses. The FIR windows
maintain the LO compartments at ambient pressure. The total mass of
the main structure, including the cryostats, is between 400 and
500$\,$kg, depending on the actual instrument configuration.

Parts of the IF processor and the digital spectrometer back-ends are
located in the co-rotating counterweight rack. The so-called PI-rack
mounted to the aircraft cabin floor holds the main IF-processor, the
acusto-optical spectrometer (AOS) and the chirp transform spectrometer
(CTS) back-ends, the power distribution, and the observing computer
with ethernet hub.

All GREAT channels flown during Early Science used the same type of
liquid nitrogen and liquid helium-cooled cryostat with a similar
wiring scheme. Special attention was given to potential safety
hazards, and detailed structural analysis was performed. Eight
identical units were manufactured so far (Cryovac, Germany), including
those assigned for the high-frequency channel. Hold times of 20+ hours
are adequate for science flights of not more than 10-11 hours duration
- typically a last top-up is performed four hours before take-off.

These channels all use waveguide-coupled hot-electron bolometers (HEB)
as mixing elements, manufactured by KOSMA
\citep[see][]{2012A&A_specialv_karl}. Devices are operated at
$<4.5\,$K, followed by a cryogenic amplifier (CITLF4, CalTech) without
intermediate isolator. All channels work in double-sideband mode.  For
the calibration a signal-to-image band ratio of 1 is used.
Solid-state cascading multiplier chains from Virginia Diodes, Inc.,
serve as LO reference \citep{2011crowe}, with commercial synthesizers
in the $10\,-\,20\,$GHz frequency range as driver stages. All channels
operate a motor-driven Martin-Puplett interferometer as LO diplexer.
The LO-power is adjusted via a rotatable wire grid in front of the
diplexer.

\subsection{Quasi-optical design}
GREAT as a dual-color receiver records two frequencies simultaneously
(the two beams must co-align on the sky within a fraction of their
beam sizes). Except for mirror curvatures, the two L-band channels
share a common design consisting of an ambient temperature optics
bench and a monolithic cold optics unit inside the cryostat
\citep{2007PhDT.......210W}. The signal from the telescope is
separated into the two frequency channels by means of a beam splitter
(polarizer grid).  The split beams propagate sideways toward "their"
optics plate carrying the Martin-Puplett diplexer for LO combining and
a focusing off-axis ellipsoidal mirror behind that defines a second
beam waist close to the plane of the cryostat window. The LO beam
enters from the bottom of the optics compartment and is redirected by
a plane mirror through an LO-attenuator toward the diplexer. Inside
the cryostat two more off-axis ellipsoids combined with a flat mirror
match the beam to the waist of the corrugated mixer feed horn.  The
Gaussian optics of the L-channels is designed for a lowest frequency
of operation of 1.2$\,$THz (here, optical elements are still sized for
beam contours of 5.5 times the waist size.)

In view of the tighter tolerances required for operation of the
M-channel, the optics layout was changed. Following the beam from the
beam-splitter, it first is re-imaged by an ellipsoidal mirror creating
a broad intermediate waist of $4.4\,$mm at the beam-splitter of the
following Martin-Puplett interferometer. The large intermediate waist
size and the now horizontal orientation of the rooftop mirrors make
the design more robust against vibrations and alignment errors. The
cold optics only consists of one off-axis mirror, matching the beam
directly to the horn antenna of the mixer.

For all channels, the wavelength dependence of the diplexer requires
mechanical adjustment of its adjustable stage to sub-$\mu$m
positioning accuracy (see \ref{DIPLEXER}, for the implementation).

The L1 channel uses high-density polyethelene (HDPE) windows for the
cryostat and the LO window in the optics compartment. Since absorption
of the material increases with frequency, we chose anti-reflection
grooved silicon windows for the L2- and the M-band channels
\citep{WagnerGentner2006249}.

\subsection{IF processors and spectrometers}
The GREAT IF system operates two IF-processors with remote-control
attenuators and total power detectors, one for the AOS and the CTS,
one for the fast-Fourier transform (FFT) spectrometer. They shift the
IF band (receiver output) to the input band of the individual
spectrometer.  Both IFs are equipped with remot-control synthesizers
for the internal mixing processes, allowing the position of individual
spectrometer bands within the IF band to be adjusted.

GREAT offers different types of back-end spectrometers that provide
different resolutions and bandwidths. The wide band AOS offers
2$\times$ 4$\times$1$\,$GHz of bandwidth with a resolution (equivalent
noise bandwidth, ENBW) of $\sim\,1.6\,$MHz
\citep{1999ExA.....9...17H}. The IF-processor is capable of stacking
four AOS-bands, thus forming two $4\,$GHz wide bands. The two CTS,
however, only have $220\,$MHz of bandwidth but a spectral resolution
of $56\,$kHz \citep{2004ExA....18...77V}. The fast advancements in
digital signal processing during recent years made it possible to
equip GREAT with FFT spectrometers \citep{2006A&A...454L..29K,
  2012A&A_specialv_bernd}: each detector channel is serviced by an
AFFT spectrometer with $1.5\,$GHz instantaneous bandwidth and
resolution of $212\,$kHz (ENBW). In June 2011 the latest technology
XFFTS was added to the system, providing $2.5\,$GHz of bandwidth with
a resolution of $88\,$kHz. For the 2012 observing season, the next
XFFTS generation with $64\,$k channels and hence $44\,$kHz spectral
resolution will be available. All back-ends can be operated in
parallel.

\subsection{Calibration unit}
\begin{figure}[b]
  \begin{center}
    \includegraphics[height=9.0cm, angle=-90]{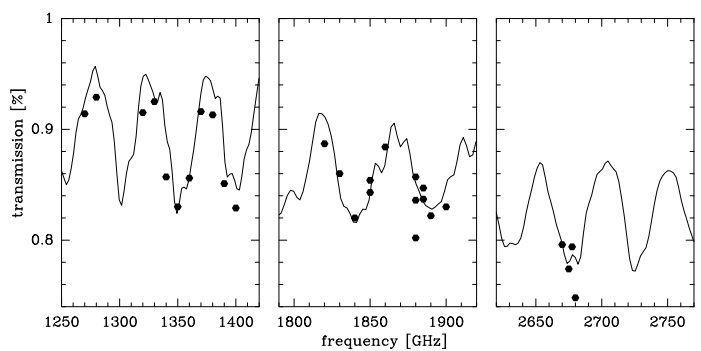}
  \end{center}
  \caption{\label{GREAT_cal_window}FTS measurements of the cold load
    window transmission (solid line) and cross calibrations (dots)
    using an external liquid nitrogen-cooled absorber without window.
    The two measurements agree well and give confidence in the
    effective cold load temperature determination, which is important
    for calibrating the science data accuratly.}
\end{figure}

Because frequent receiver gain calibration is essential for
high-quality astronomical data, GREAT hosts an internal calibration
unit with two black body radiators, one at liquid nitrogen (cold load)
and one at ambient temperature (hot load). The cold load cryostat has
a hold time of $\,>15\,$hrs. A sliding mechanism close to the
telescope's focal plane re-directs the beam to the cold load, to the
hot load, or keeps the aperture clear for the sky signal. Depending on
the actual observing condition, reference load measurements are taken
every 5-10 minutes.

Proper calibration of losses through the residual atmosphere at flight
altitude requires accurate knowledge of the effective load
temperatures \citep[see][]{2012A&A_specialv_juergen}.  Therefore we
carefully measured the transmission of the HDPE window of the cold
load cryostat with a Fourier transform spectrometer and
cross-calibrated these measurements against an external windowless
liquid nitrogen cooled load (Fig. \ref{GREAT_cal_window}). With the
window transmission known, the effective cold load temperature is then
calculated individually for each receiver sideband from the physical
load temperature measured by three temperature sensors on the cold
absorber cone. The overall temperature uncertainty, dominated by the
inaccuracy of the window transmission curve ($<\,5\,$\%), is
$<\,5\,$K.  This corresponds to an approximately $2.5\,$\% intensity
calibration uncertainty -- much lower then the uncertainties in
telescope efficiency (Sect. \ref{Calib}) and in atmospheric
calibration \citep[see][]{2012A&A_specialv_juergen}.

\subsection{GREAT receiver control}
GREAT uses two independent Linux-based computers to control the
front-end and to run the observing software, a server for the
observing software, and a VME-system for the main tasks of the
front-end control.

The optics control electronics provides the calibration unit controls
(drive mechanism, temperature readout) and the two controllers for the
diplexer drive systems. This system is computer-controlled via RS232
serial lines. The four-channel bias supply for the HEB detectors and
the bias supply for the cold low-noise amplifiers together with the
cryostat temperature readout are directly connected to the
VME-computer I/O hardware, giving full remote control of the
front-end.

\subsubsection{Software structure}
The software that controls the hardware reflects the modularity of the
hardware: to each hardware module that can be software-controlled, an
individual software task (server) is assigned.  In addition virtual
devices exist for modules without direct hardware counterpart.
Toghether, these modules provide all necessary functions to operate
GREAT. The individual software tasks are spread within the GREAT
computer network and communicate via the UDP-protocol. The full
software package can be configured by human readable ASCII text files,
which also denote which task is running where. All servers use the
same code basis and differ only in their configuration files and the
hardware-specific code. A hardware simulator to test the higher level
software was realized by exchanging the hardware-specific parts only.

All user interfaces, including the graphical user interface, have been
realized in Perl/Tk based on a dedicated Perl macro package. The macro
package also gives easy access to the full hardware functions within a
script-based language. This allows the user to quickly develop special
purpose tasks, e.g., for debugging.

\subsubsection{Automatic tuning}
Since observing time on an airborne telescope is precious, we have
automated the routine operations during observations as much as
possible. To change observing frequencies, multiple systems have to be
adjusted and re-tuned. For the current GREAT channels we need to set
the LO frequency, adjust the LO power level, tune the interferometric
diplexer, and level the IF-chain output power. The HEB mixer can be
operated on a fixed bias point for all frequencies. Different bias
points for the diplexer optimization and for the LO-power adjustment
can be chosen.

\paragraph{LO tuning:}
The algorithm automatically changes the tuning frequency (following
pre-defined procedures to safeguard the hardware) and adjusts its
power by measuring the mixer BIAS current at a given point. The
optimization is a Newton search using tabulated start values. The L2
channel uses a first incarnation of an electronic LO-power
stabilization system. The actual mixer current is kept constant by
changing the LO-chain driver amplifier gain. This system also
compensates for the direct detection effect (see also below).

\begin{figure}[ht]
  \begin{center}
    \includegraphics[height=4.5cm,
    angle=-90]{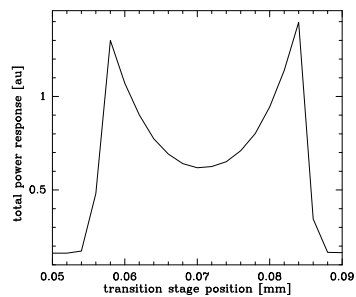}\hfill
    \begin{minipage}[t]{4.0cm}
      \vspace{-2ex}
      \caption{\label{GREAT_diplexer} Total power response of the HEB
        mixer versus diplexer roof top position. The shape of the
        curve results from decreasing mixer gain with increasing LO
        pump power \citep[see][for details]{2012A&A_specialv_karl}.
        Noise temperature and stability of the system are best at the
        response minimum.}
    \end{minipage}
  \end{center}
  \vspace{-6ex}
\end{figure}

\paragraph{Diplexer optimization:\label{DIPLEXER}}
For LO coupling we use Martin-Puplett interferometers in all three
channels that are currently in operation. The interferometer is
pre-adjusted by calculating a model fit to -- previously measured --
grid points. This is, if set up well, accurate to better than
$5\,\mu$m, but an optimum injection of the THz radiation for best
noise temperature requires positioning the interferometer to $500\,$nm
accuracy. This final adjustment is made, again relying on a Newton
search, by minimizing the total power response (Figure
\ref{GREAT_diplexer}) of the mixer. Setting the LO and tuning the
diplexer takes 1-2 min per channel.

\paragraph{Compensating direct detection:}
HEB mixers are known to respond to direct detection
\citep[see][]{2012A&A_specialv_karl}, which is the change of the mixer
current (and gain) in response to the load by the incoming radiation.
Because the effect becomes more prominent with decreasing size of the
device, we implemented a correction scheme for the $2.5\,$THz HEB
(this channel showed a noticeable direct-detection effect on the order
of 50\% in noise temperature, while the lower channels operate with
larger devices and showed no measurable effect): after switching to
the loads or back to sky we re-adjust the LO power with the optical LO
attenuator, and thereby operate the mixer at constant current
throughout.

\begin{figure*}[bt]
  \begin{center}
    \includegraphics[height=9cm,angle=-90]{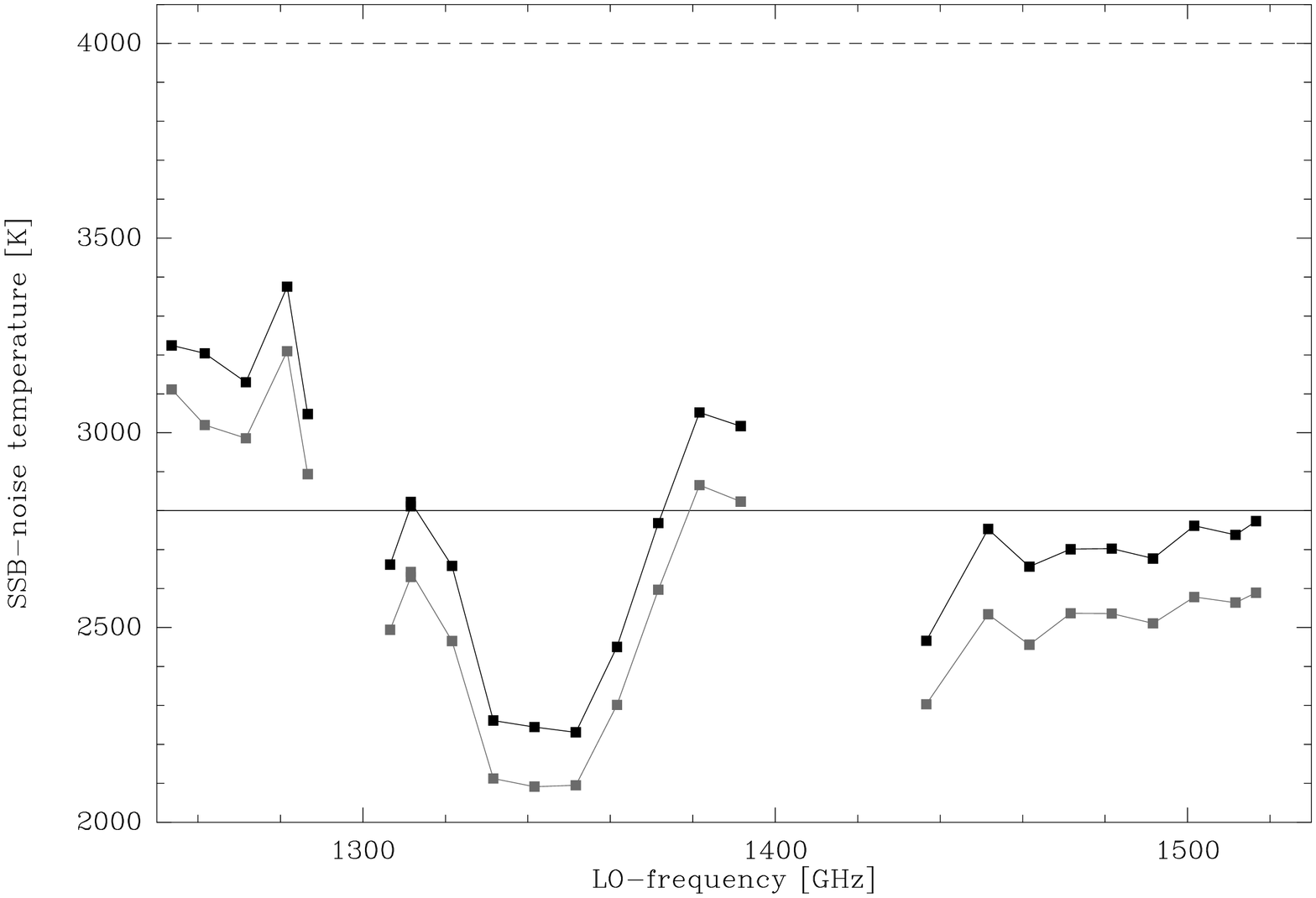}\hfill
    \includegraphics[height=9cm,angle=-90]{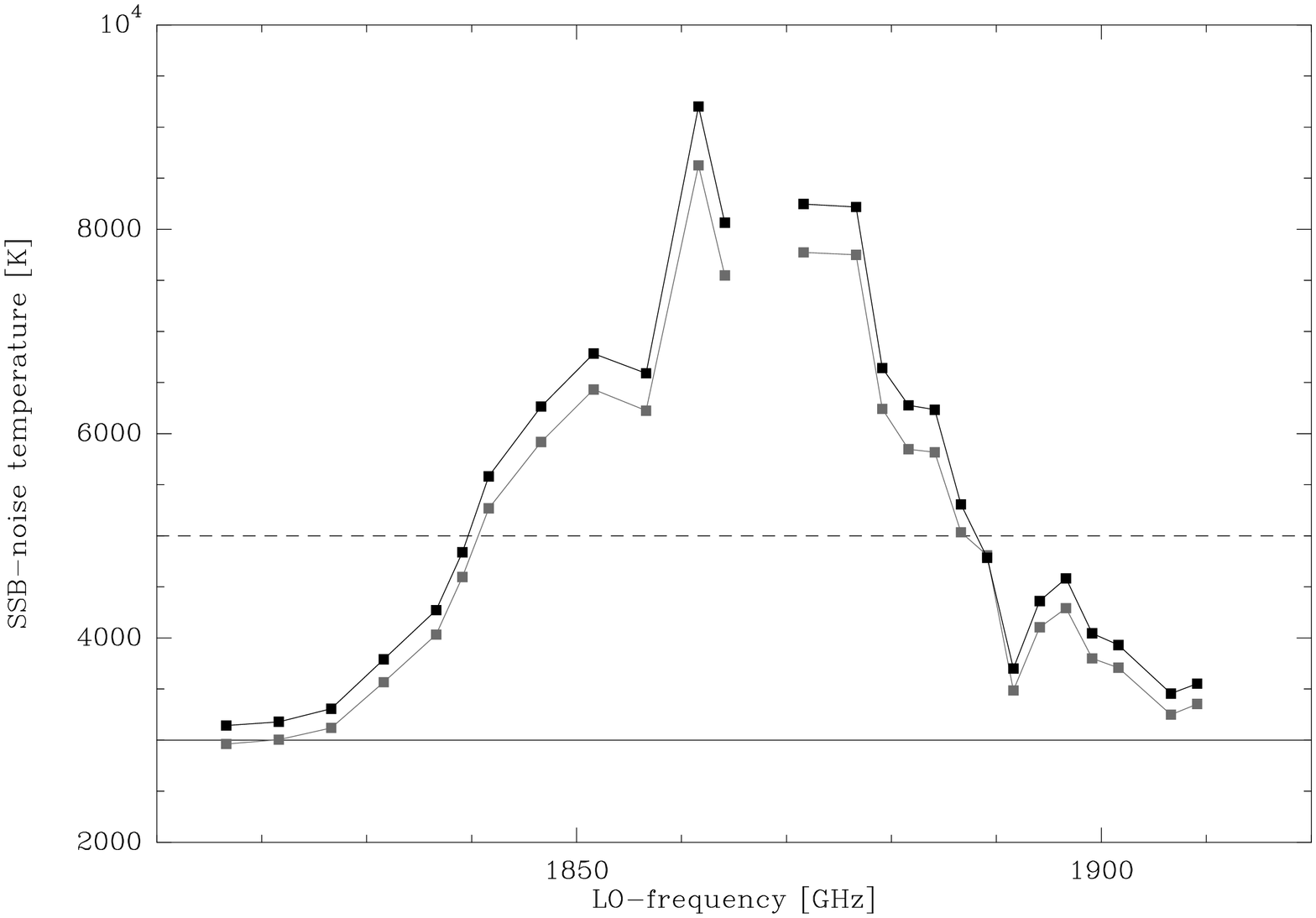}
  \end{center}
  \caption{\label{GREAT_noise_low} Single-sideband receiver noise
    temperatures vs. tuning range for bands L1 (left panel) and L2
    (right panel). Horizontal lines denote the baseline (dashed lines)
    and goal performance committed for Early Science. The two curves
    display average noise temperatures - around the IF-center - for IF
    bandwidths of 1.0 and $1.5\,$GHz (higher noise).}
\end{figure*}

\section{Instrument alignment}
Rare and expensive flight time together with an estimated turn-around
time for optics re-adjustments during flight of $>\,1\,$hour, make it
mandatory to ensure proper alignment before flight.  The co-alignment
between the channels and the optical axis of the telescope is to be
established by pre-flight alignment procedures -- in the GREAT
laboratory at the DAOF and, after installation, in the aircraft.
Because of the high athmospheric absorption at sea level, GREAT is
blind on sky unless at high flight altitudes, and the determination of
the "boresight" offsets to the optical guide cameras can only be
performed in flight (once established, the positioning of GREAT toward
the astronomical target is defined and maintained by the optical
cameras).

In the laboratory, both receiver beams are adjusted to be centered on,
and perpendicular to, the instrument mounting flange (consequently
also co-aligned). Two beam scanners are used to characterize the beams
at two locations, one inside the optics compartment, near the
telescope's focal plane (see Fig. \ref{GREAT_optics}), and one at a
distance of several meters away from the flange. The reference
positions for the beam scanners are defined with a visible laser. Once
inside the aircraft, we check that the two beams are well-centered on
the secondary mirror by monitoring the total power response of the
detectors against cold absorber paddles, but we do not touch their
internal alignment. In a last step in flight, the boresight offsets to
the guide cameras of SOFIA are determined with pointing scans across a
bright planet (that is visible to GREAT and the guide camera, see Fig.
\ref{GREAT_Beam}).  The overall pointing accuracy therefore depends on
the accuracy of the boresight determination (typically 1 - 2$\,$") and
the actual co-alignment between the channels (Sect. \ref{Calib}). The
stability of the pointing during science operation is controlled with
the optical guide cameras to 3 - 5$\,$" (in a trade-off between
science target integration efficiency and overhead for optical
peak-up).

\section{GREAT performance during Early Science}
GREAT participated in 19 flights, including observatory
characterization, commissioning and transfer flights to Europe, during
four installation periods. Seven flights (including three for Short
Science) were reserved for use by the GREAT consortium, 11 flights
assigned for Basic Science community projects on a shared-risk basis;
as expected for a new observatory and instrument, not all were
successful. Nevertheless, including science data taken during the
commissioning flights, a total of 25 science projects were
successfully executed.

\begin{figure}[b]
  \begin{center}
    \includegraphics[height=9cm,angle=-90]{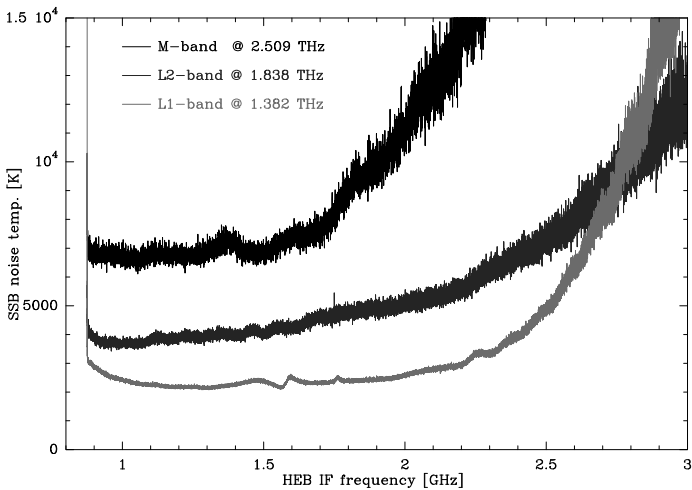}
  \end{center}
  \caption{\label{GREAT_bw} Typical variation of SSB receiver noise
    temperatures with IF frequency for all three channels.}
\end{figure}

\begin{figure}[tb]
  \begin{center}
    \includegraphics[height=9cm,angle=-90]{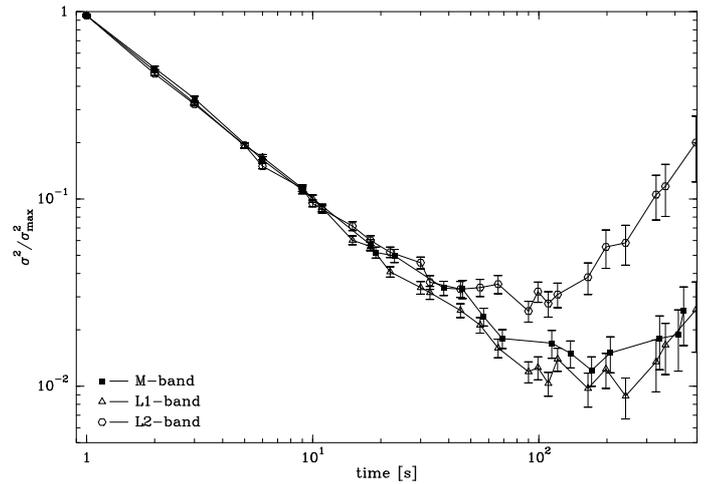}
  \end{center}
  \caption{\label{GREAT_Allan} Spectroscopic Allan variance plots for
    bands L1, L2 and M, calculated between two AFFTS channels
    separated by $750\,$MHz. The channel width was set to $850\,$kHz.}
\end{figure}

Three detector channels were operated during Early Science (see Table
\ref{GREAT_channels}) in the configurations L1$\rm_a$/L2, L1$\rm_b$/L2
and M$\rm_a$/L2. The successful commissioning of our latest
development, the M$\rm_a$ channel operating at $2.5\,$THz,
demonstrates the power of a PI instrument such as GREAT to use
state-of-the-art technologies mere weeks after they become available.
This capability is unique for SOFIA, as compared to long-planned
satellite missions. High-resolution spectroscopy beyond $1.91\,$THz,
above the bands of HIFI on Herschel, is currently only offered by
GREAT on SOFIA.

All three GREAT channels show state-of-the-art sensitivity figures
combined with a high system stability, allowing for deep integrations
and long on-the-fly (OTF) slews. With reference to performance figures
presented during the instrument's pre-shipment review ("baseline") at
many frequencies the sensitivities have improved by factor of 2, some
IF bandwidths have been tripled. The $2.5\,$THz receiver was not even
on the horizon at that time. Below we discuss the basic performance
figures.

\paragraph{Sensitivities:}
Fig. \ref{GREAT_noise_low} displays single-sideband (SSB) receiver
noise temperatures across the RF tuning ranges of the L-band channels.
While, fortunately, not impacting at the frequencies of the heavily
requested transitions of [CII, $1.90\,$THz] and [OH, $1.83\,$THz],
increased noise temperatures are recorded in the middle of the L2
band, very likely because of LO noise contributions (a replacement is
planed for the 2012 observing season). For the (narrow band) M$\rm_a$
channel the SSB noise temperature (see also Fig. \ref{GREAT_bw}) at
the OH line is $\sim 7000\,$K over the usable IF-band of $\sim
900\,$MHz.

\paragraph{IF bandwidth:}
The L-band channels are limited by the mixer roll-off at about
$3\,$GHz and the Martin-Puplett transmission band. For optimum
performance with maximized bandwidth we operated at an IF center
frequency of $1.625\,$GHz, with a diplexer center frequency of
$1.750\,$GHz: the higher diplexer frequency partly compensates for the
mixer sensitivity roll-off toward higher IF frequencies, while it
affects the sensitivity at the lower IF band only negligibly
($\sim\,$5\%). The so achieved usable bandwidth is clearly
channel-dependent: while for the L1 band a $3\,$dB noise bandwidth of
$1.8\,$GHz is attainable, the M-band has a bandwidth of $0.9\,$GHz
only (compare Fig.\ref{GREAT_bw}). The reason for the limitation in
the M-band channel is still under investigation.

\paragraph{System stability:}
Excellent spectroscopic Allan variance minimum times of up to $100\,$s
and longer were measured for the integrated system, while looking at
the hot load (Fig. \ref{GREAT_Allan}). This is excellent for a FIR
spectrometer based on HEB detectors, and is the reason for GREAT's
ability to perform long on-the-fly observations with phase times of
several 10 seconds.

\subsection{Observing with GREAT}
\begin{figure}[b]
  \begin{center}
    \includegraphics[width=9cm,angle=0]{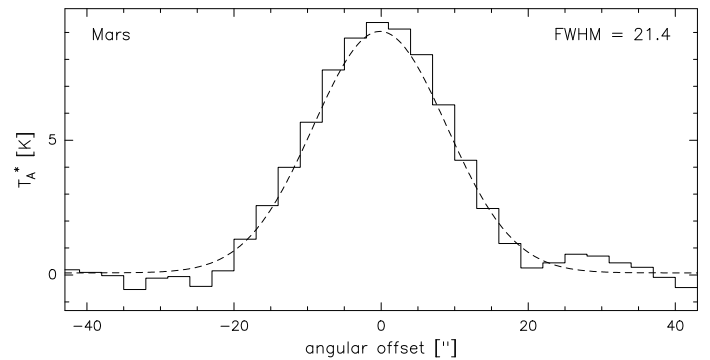}
  \end{center}
  \caption{ \label{GREAT_Beam} Total power scan of GREAT's L1-channel
    tuned to $1.337\,$THz across Mars. The measured beam is
    diffraction-limited (fit result shown as dashed line).}
\end{figure}

The observing modes supported by GREAT are (1) classical position
switching (moving the telescope), (2) beam-switched observations
(chopping with the secondary, at a rate of 1 -- 2$\,$Hz), for
single-position or raster mapping, and (3) OTF scanning (data are
collected while the telescope is slewing across the source). In view
of the good stability of the instrument, the OTF mode is most
efficient for extended structures, while beam-switching is advised for
the observation of weak(er), broad(er) lines toward more compact
objects. The wide throw of several arcminutes, with selectable chop
direction, is a most valuable asset of the SOFIA chopper.

All Early Science observations were carried out via observing scripts
in the environment of the KOSMA-developed {\it kosma\_control}
observing software. For thorough pre-flight testing of the scripts, a
realistic stand-alone software simulator is part of this observing
package. This simulator was also essential for instrument- and/or
telescope-independent software interface testing (toward e.g. the {\it
  SOFIA Mission Control and Command Software}, the front-end or the
back-ends).

Almost in real-time, the on-line calibration pipeline displays the
science data, which is valuable for real-time decisions. Post-flight,
GREAT delivers raw data in modified FITS format into the SOFIA
archive, and finally calibrated data \citep[see
also][]{2012A&A_specialv_juergen} in standard CLASS format, which was
adapted to include a SOFIA specific header section.

\section{In-flight calibration and efficiencies\label{Calib}}
By drift scans across bright planets (Saturn in Spring, Jupiter during
the Summer flights) we verified the {\it{on-sky}} co-alignment between
the two channels that were actually flown, and determined the (common)
focus of the instrument. Observations of Mars (size $5.1\,$" on Sept.
30) yielded a de-convolved half-power beam width of 21.1" $\pm\,1\,$"
at $1.337\,$THz (this frequency is free of terrestrial and Martian
atmospheric spectral features), very close to the predicted $21.3\,$"
for our diffraction-limited optics with a $14\,$dB edge taper at the
subreflector (see Fig. \ref{GREAT_Beam}). For all flights the pointing
of the two channels was co-aligned to better than $4\,$".

The T$_A^*$ calibration scale of GREAT is based on frequent (typically
every $10\,$min) measurements of our integrated hot-cold reference
loads. Corrections against atmospheric absorption are performed with
the {\it kalibrate}-task \citep{2012A&A_specialv_juergen} as part of
the observing software by spectral fits to the measured sky
temperatures across the reception band. The final step to source
coupled temperatures, T$_{mb}$, requires knowledge of the main-beam
coupling efficiencies. On Jupiter and Mars we determined $\eta_{mb}$
$\sim$0.54 (at $1.3\,$THz), 0.51 ($1.9\,$THz) and 0.58 ($2.5\,$THz),
with estimated 10$\%$ uncertainties. This comparatively low throughput
is mainly due to the large aperture blockage by the tertiary and the
oversized scatter cone at the subreflector. Accumulating only these
quantifiable losses (plus blockage through the spiders and the
dichroic and reflection losses) we calculated upper limits to the beam
efficiencies of 0.56, 0.57 and 0.59, respectively - only slightly
higher than what has actually been observed. The differences are
attributed to "unspecified losses" such as spill-over, wavefront
errors, or optical imperfections (misalignments, incorrect foci).

\begin{table}[t]
  \caption{\label{GREAT_performance}GREAT performance (during Basic Science)}
  \begin{center}
    \begin{tabular}{|l|c|c|c|c|}
      \hline
      Channel    & RF tuning    & T$_{rec}$(SSB) & $\Theta_{mb}$ & $\eta_{mb}$  \\
                 &   [GHz]      &  [K]           &  [arcsec]     &              \\\hline\hline
      L1$\rm_a$  & 1252 - 1392  & 2000 - 3500    &   21.3        & $\sim0.54$   \\\hline
      L1$\rm_b$  & 1417 - 1520  & 2500           &   19.6        & $\sim0.54$   \\\hline
      L2         & 1815 - 1910  & 3000 - 8000    &   15.0        & $\sim0.51$   \\\hline
      M$\rm_a$   & 2507 - 2514  & 7000           &   11.4        & $\sim0.58$   \\\hline
    \end{tabular}
  \end{center}
  The RF tuning range has narrow gaps
  from 1285 - 1312 and 1860 - 1875$\,$GHz. Receiver noise
  temperatures are SSB. The main beam FWHP for diffraction-limited
  optics was verified by measurements of Mars with band L1$\rm_a$. Allan
  minimum times as shown in fig. \ref{GREAT_Allan} are above $80\,$s for
  all three channels.
\end{table}

\section{Conclusions and outlook}
With GREAT a state-of-the-art dual-color heterodyne receiver for
operation onboard of SOFIA has been developed that already during its
first observing cycles provided fascinating new insights into the
far-infrared universe. The scientific impact is already clearly
evident by the large number of scientific contributions to this A$\&$A
special volume, which is dedicated to Early Science with GREAT/SOFIA.

The consortium intends to maintain GREAT at the leading edge of THz
technologies. GREAT will be continuously upgraded with new
opportunities as they become available, improving the performance of
existing frequency channels and adding new bands -- thereby providing
better sensitivities, increasingly wider RF coverage and IF bandwidth
to our users.  In preparation for Cycle 1, new LO sources will be
integrated (band L2, M), including our novel photonic local
oscillators, and improved HEB devices will be manufactured (L2, M).
New frequency channels (M$\rm_b$, H) will become available and the FFT
spectrometers will then be upgraded to $64\,$k channels ($44\,$kHz
resolution). Once successfully commissioned, the improved GREAT will
be available again to the interested SOFIA communities in
collaboration with the consortium (following conditions similar to
those implemented for Basic Science).

Since last summer upGREAT is on course -- a parallel development to
the improvements of the existing channels that building on the
infrastructure of GREAT, aims at delivering two compact heterodyne
detector arrays with $2\times7$ pixels (1.9-2.5$\,$THz) and $1\times7$
pixel ($4.7\,$THz), respectively. Within an ambitious schedule, we aim
at commissioning the first of the low-frequency sub-arrays in Cycle 2.

\begin{acknowledgements}
  During its long development many have contributed to the success of
  GREAT, too many to be named here. We thank the SOFIA engineering and
  operations teams, whose tireless support and good-spirit teamwork
  has been essential for the GREAT accomplishments during Early
  Science. Herzlichen Dank to the DSI telescope engineering team.
  Finally, we thank the Project Management for bringing SOFIA back on
  schedule!

  IRAM/GRENOBLE has implemented the GREAT/SOFIA specific CLASS header
  extensions.

  SOFIA Science Mission Operations are conducted jointly by the
  Universities Space Research Association, Inc., under NASA contract
  NAS2-97001, and the Deutsches SOFIA Institut under DLR contract 50
  OK 0901.

  The development of GREAT was financed by the participating
  institutes, the Max-Planck-Society, and the German Research Society
  within the framework of SFB 494.

\end{acknowledgements}
\bibliographystyle{aa}
\bibliography{GREAT_aa}
\end{document}